# A Novel Experimental Approach for Nanostructure Analysis: Simultaneous Small Angle X-ray and Neutron Scattering (SAXS/SANS)

**Ezzeldin Metwalli[a], Klaus Götz[ab], Sebastian Lages[a1], Christian Bär[a], Tobias Zech[ab], Dennis M. Noll[a], Isabel Schuldes[a], Torben Schindler[a], Annemarie Prihoda[ab], Herbert Lang[a], Jürgen Grasser[a], Mark Jacques[c], Luc Didier[c], Amrouni Cyril[c], Anne Martel[c], Lionel Porcar[c] and Tobias Unruh[ab]***

[a]Institute for Crystallography and Structural Physics, Friedrich-Alexander-Universität Erlangen-Nürnberg, Staudtstr. 3, 91058, Germany
[b]Center for Nanoanalysis and Electron Microscopy (CENEM) and Interdisciplinary Center for Nanostructured Films (IZNF), Friedrich-Alexander-Universität Erlangen-Nürnberg (FAU), Cauerstr. 3, Erlangen, 91058, Germany
[c] Institut Laue-Langevin, 71, Avenue des Martyrs, Grenoble, 38042, France

*Correspondence email: tobias.unruh@fau.de
[1]Current address: Department of Physics, Chalmers University of Technology, SE-412 96 Göteborg, Sweden

**Synopsis** Portable SAXS instrument with proper geometrical dimensions for installation at D22 instrument was designed and constructed for simultaneous SAXS/SANS experiments at ILL.

**Abstract** Exploiting small angle X-ray and neutron scattering (SAXS/SANS) on the same sample volume at the same time provides complementary nanoscale structural information at two different contrast situations. Compared with an independent experimental approach, the truly combined SAXS/SANS experimental approach ensures the exactness of the probed samples particularly for in-situ studies. Here, we introduce an advanced portable SAXS system that is dimensionally suitable for installation at D22 zone of ILL. The SAXS apparatus is based on a RIGAKU copper/molybdenum switchable microfocus rotating anode X-ray generator and a DECTRIS detector with a changeable sample-to-detector distance of up to 1.6 m in a vacuum chamber. A science case has been presented to demonstrate the uniqueness of the newly established method at ILL. Temporal structural rearrangements of both, organic stabilizing agents and organically capped gold colloidal particles during gold nanoparticle growth are simultaneously probed, enabling immediate correlated structural information. The newly established nano-analytical method at ILL will open the way for real time investigations of a wide range of innovative nanomaterials and will enable comprehensive in situ studies on biological systems. A potential development of a fully-automated SAXS/SANS system with a common control environment and additional sample environments, permitting a continual and efficient operation of the system at the hands of ILL users, has also been introduced.

## 1. Introduction

Small angle scattering (SAS) of X-rays (SAXS) and neutrons (SANS) is a powerful experimental technique that is widely used in materials science, soft condensed matter, and structural biology (Fitter *et al.*, 2006, Glatter & Kratky, 1982, Svergun *et al.*, 1987, Svergun *et al.*, 2013). Both the SAXS and SANS methods provide statistically relevant structural information on length scales from 0.1 to 100 nm at two contrast situations (Blanchet &

Svergun, 2013, Koch *et al.*, 2001, Svergun, 2010). The X-ray interacts prevalently with the electron shell of the atoms, while neutrons interact with their nuclei, thus, both radiations offer two different contrasts (X-ray and neutron scattering cross-section). The SAS method has been used with great success in investigating the structure of soft condensed matter such as emulsions (Schmiele *et al.*, 2016), micelles (Schmutzler, Schindler, Goetz*, et al.*, 2018, Schmutzler, Schindler, Schmiele*, et al.*, 2018), liquid crystalline structures (Gehrer *et al.*, 2014, Schmiele, Gehrer*, et al.*, 2014, Schonhals *et al.*, 2010), and organic nanoparticle (NP) dispersions (Schuldes *et al.*, 2019, Schmiele, Schindler*, et al.*, 2014, Unruh, 2007). For advanced functional materials (e.g., sensors, solar cells, and lithium ion batteries), SAXS and SANS are important techniques to study the formation, growth, and stabilization of inorganic nanomaterials (Schindler *et al.*, 2015, Schindler *et al.*, 2017, Wang *et al.*, 2015, Wang *et al.*, 2019, Yin *et al.*, 2018, Zheng *et al.*, 2018, Mohl *et al.*, 2018, Futscher *et al.*, 2019, Metwalli *et al.*, 2015). In modern biological applications, the SAS method is an essential tool for the structural characterization of proteins, nucleic acids, and lipids, as well as for monitoring structural changes during protein folding, intrinsic disorder, conformational transitions, and protein-protein assembling processes (Fitter *et al.*, 2006, Svergun *et al.*, 2013, Blanchet & Svergun, 2013, Koch *et al.*, 2001, Schindler *et al.*, 2018). In contrast to analytical methods such as scanning or transmission electron microscopy (SEM/TEM), SAS offers a non-destructive bulk characterization technique allowing analyses of a small sample volume at very low concentrations (< 10 mg/mL) and a rapid data acquisition capability (Svergun *et al.*, 2013). In other words, the dispersed materials can be easily and reliably characterized in a solution using SAS without extensive preparation such as drying or plunge freezing (Frank, 2002), in contrast to other analytical methods where potential artifacts prior to characterization are likely to occur. By employing proper software tools that enable high throughput data reduction and efficient data analysis, time-resolved SAS measurements can uniquely be utilized for studies on growth rates and phase transitions of colloidal dispersions in real-time mode (Graceffa *et al.*, 2013, Jensen *et al.*, 2010, Tanaka *et al.*, 2007, Arleth *et al.*, 2014, Illing & Unruh, 2004). Thus, SAS holds great promise for answering many unsolved scientific questions, especially for soft matter and structural biology research.

## 1.1. SAXS and SANS: Contrast variations and wide angle option

Both SAXS and SANS provide global shape, size and size distributions as well as spatial distribution of dispersed macromolecules in solution (Putnam *et al.*, 2007). In particular, they provide the diameter and mass of the gyration of isotropic-shaped particles, as well as the linear mass density and cross-sectional size of anisotropic-shaped particles (Putnam *et al.*, 2007, Koch *et al.*, 2003). The internal structure of many biological systems and colloidal dispersions is complex and complementary to the global structural details (Whitten, Jeffries*, et al.*, 2008, Grossmann *et al.*, 2008). While, the SAXS method is sensitive to the electron-density difference between dispersed nano-sized moieties and their environment, the SANS method uses neutrons as probes which are sensitive to different isotopes of the same element allowing variation of contrasts and internal structural determination. Thus, both SAXS and SANS provide complementary information about the structure of the investigated samples. There are many aspects to the complementarity of SAXS and SANS (Schmidt, 1995, Zemb & Diat, 2010). One important aspect is the contrast variation. For instance, the contrast-variation SAXS has been used to investigate the internal electron density structure of a system by employing chemical agents such as glycerol, sucrose, gadolinium-based molecule, and salt (Grishaev *et al.*, 2012, Chen *et al.*, 2014, Kuwamoto *et al.*, 2004). Due to the potential chemical and physical modifications of the samples upon incorporating these agents and their limited electron density ranges, the utility of SAXS-based contrast-variation measurements is not substantial. Additionally, the viscosity of a liquid sample was reported to increase after using a high concentration of contrast agents such as sucrose (Chen *et al.*, 2014). In a viscous system, a potential effect on a transition from a non-equilibrated to an equilibrated state may affect the evolved nanoscale structure. Furthermore, the applicability of high salt concentrations for studying biological systems in contrast-variation SAXS is narrow due to undesired chemical modifications (Chen *et al.*, 2014). For neutrons, the contrast-variation option is, however, more popular and can be achieved by targeted isotopic substitution of the nanostructured components or by changing the ratio of deuterated/protonated solvent composition of the investigated samples (Neylon, 2008, Sugiyama *et al.*, 2016). Using contrast-variation SANS, the resulting scattering profile of a particular component and its relative orientation in a multicomponent assembly can be retrieved (Whitten, Cai, *et al.*, 2008, Schmutzler, Schindler, Schmiele, *et al.*, 2018). This would reveal molecular recognition and the unique relative positioning of the components within the nanoscale assemblies (Svergun & Nierhaus, 2000). In other words, contrast-variation SANS is a uniquely suited method for probing the structure of a particular component in the presence of another, by employing a contrast-matching experimental approach. The wide angle option of the neutron-scattering method is limited in investigating atomic and small-sized structure samples due to the broad wavelength band ($\Delta\lambda/\lambda \cong 10\ \%$) that is typically employed for SANS instruments. A high resolution ($\Delta\lambda/\lambda \cong 1\ \%$) SANS instrument at National Institute of Standards and Technology (VSANS-NIST) is available for investigating a large q range (molecular resolution) but at a highly reduced flux. Alternatively, X-ray scattering reliably enables a wide angle scattering (WAXS) option, which satisfactorily provides information on the atomic resolution and finer structural features of the investigated systems.

**1.2. Combined SAXS and SANS: Independent measurements**

From the above discussion it is obvious that both SAXS and SANS are truly complementary techniques that enable a wide q dynamic range for structural information—from atomic and molecular scales to large assemblies. Despite the SANS technique's unique advantage in internal structural determination (Schindler *et al.*, 2015), many multicomponent entities in an assembly continue to pose a considerable challenge to neutrons. Thus, the SAXS technique is used to obtain comprehensive structural details. For many science cases (e.g., core-shell colloids, biological systems, nanoparticle-protein complexes), the need to combine SAXS and SANS to resolve structural information has been reported (Schindler *et al.*, 2015, Hennig *et al.*, 2013, Spinozzi *et al.*, 2017, Schindler *et al.*, 2018, Schuldes *et al.*, 2019). As an example, for the suspension of organic-coated solid inorganic NPs (Schindler *et al.*, 2019, Schindler *et al.*, 2015, Schindler *et al.*, 2017), the SANS method is particularly sensitive to the hydrogenated organic shell while the core is transparent to neutrons. Thus, SANS will provide the form factor of the shell in a representative hollow-shell-like particle (Schindler *et al.*, 2019). In contrast, SAXS measurements of the same core-shell sample are sensitive to the high electron density core and significantly less sensitive to the shell. Both, shell and core structures can thus only be obtained via complementary SAXS and SANS methods. For organic core-shell systems, the opposite scenario is true. For systems such as surfactant (sodium dodecyl sulphate; SDS)-coated cetyltrimethylammonium bromide (CTAB) micelles, the shell is invisible to the SANS method due to enrichments of employed $D_2O$ in the shell (Bergstrom & Pedersen, 2000). The form and structure factors of the hydrogenated core can thus be obtained. For this organic core-shell sample, the SAXS method, however, will provide structural information of the shell-like structure due to high electron density counter-ions in the electrical double layer within the shell. Thus, both SAXS and SANS are complementary methods for detailed structural information of many systems, especially for in-situ studies. For biological systems, combined SAXS/SANS investigations have become an established tool to get a complete structural picture together with computer-based structure calculation protocols. For instance, contrast-variation SANS investigation of subunit-selectively deuterated samples can provide valuable additional information on these subunits, such as the protein-RNA complexes (Hennig *et al.*, 2013). Ambiguities in the resulting analysis structural models of the SANS data can be further resolved by a SAXS experiment, which should yield information about the overall shape of the complex molecular structure (Hennig *et al.*, 2013). The importance of joint SAXS/SANS experiments has already been recognized. For instance, an initiative (SAS platform) at large scale facilities (ILL and ESRF) have been launched, allowing joint beamtime allocation on both SAXS (BM29-ESRF) and SANS (D22-ILL) instruments (Lapinaite *et al.*, 2013).

**1.3. Combined SAXS and SANS: Simultaneous mode**

An analytical concept based on coupling both SAXS and SANS methods in a single experiment is an important step forward in the direction of novel analytical technologies that pave the way to innovations in advanced nanostructured materials. The key advantage of simultaneous SAXS/SANS measurements is the ability to probe a particular sample at the same time in two different contrast situations for X-rays and neutrons. It is almost impossible to check if two samples are really identical by performing independent SAXS and SANS measurements, especially at different contrast-variation levels (deuteration and buffers). Thus, simultaneous SAXS/SANS measurements allow both the exactness of the probed samples and the potential of using simultaneous structural analysis models for both SAXS and SANS data, consequently presenting a straightforward and unambiguous interpretation of the scattering data. A key contribution of the simultaneous SAXS/SANS method is its ability to perform in-situ and operando studies on materials that undergo modifications upon relevant environments, allowing access to their temporal structural rearrangements from molecular scale to nanoscale assemblies at two different contrast situations.

In this work, we will introduce the technical description of a custom-made portable SAXS system that is geometrically suitable for installation at the zone of a SANS instrument (D22) of ILL. The whole system's components are uniquely mounted on a single metal chassis so it can be craned and quickly moved inside the D22 zone. Within one hour, the SAXS instrument can be installed in the D22 zone, allowing a fast start of simultaneous SAXS/SANS experiments. Furthermore, we will emphasize on a lead shielding configuration that efficiently stops high energy gamma so minimum background on the X-ray detector is achieved. Both X-ray and neutron beams are orthogonal and will be superimposed on the same sample positioned at 45° with respect to both beams, allowing simultaneous SAXS/SANS measurements. We will present the full specifications of the new setup and the first science experiment using the simultaneous SAXS/SANS method. Finally, we will describe further

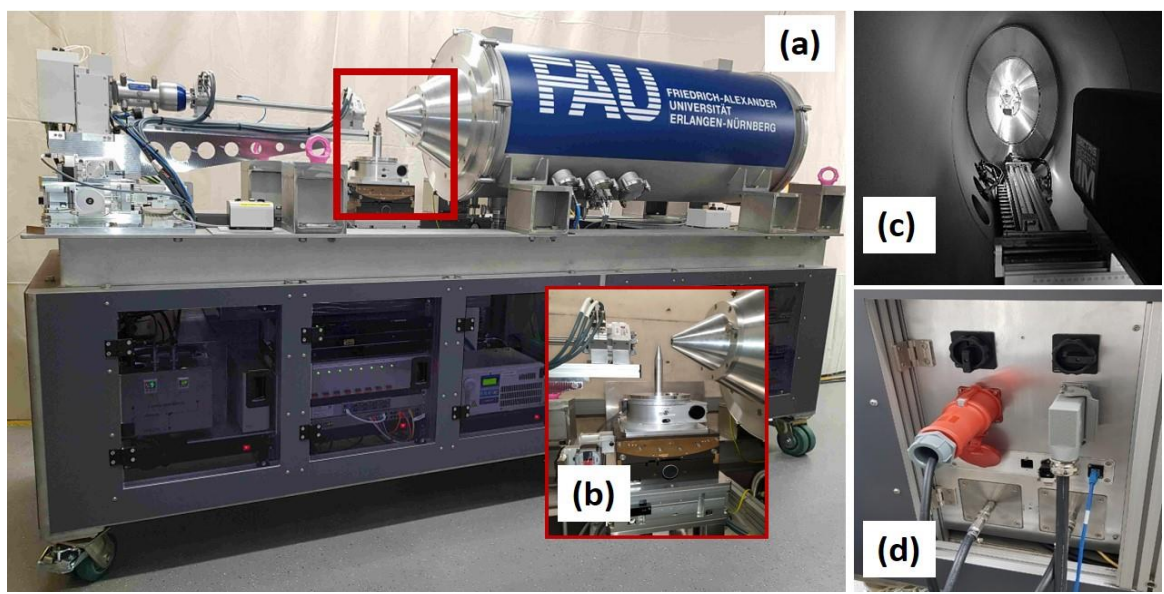

**Figure 1** a) Photograph of the portable SAXS instrument for simultaneous SAXS/SANS measurements at D22-ILL, b) 6-axis goniometer for proper sample alignments with respect to both X-ray and neutron beams, c) EIGER R 1 M detector with three degrees of freedom to move inside a vacuum chamber, d) a central hub for an easy and smooth plug-and-play operation, including power supplies, cooling-water and Ethernet cable. All system components are enclosed within a single heavy-duty metal non-magnetic chassis.

potential developments, which include further minimization of background radiation on the X-ray detector and a new control software within ILL software tools (NOMAD) to facilitate a potent approach for combining both SAXS and SANS methods at the hands of ILL users.

## 2. General description

An advanced SAXS system based on a RIGAKU copper/molybdenum switchable microfocus rotating anode X-ray generator and an EIGER R 1M detector with a changeable sample-to-detector distance (SDD) from 0.5 m up to 1.6 m in a vacuum chamber was designed and constructed (**Figure 1**). The SAXS system is dimensionally suitable for installation at the D22 instrument of ILL.

With a Rigaku VariMax™ optics and a beam divergence control system directly connected to the X-ray source, switching from one wavelength to the other (Cu $K_\alpha$ or Mo $K_\alpha$) is easily achievable. An evacuated collimation system of about 56 cm in length is composed of three fully-automated vacuum slits. Each slit has four independently moveable blades which define the aperture size and position. A tungsten carbide and two scatterless slit system equipped with scatterless Si and GaAs blades define a variable size and low divergent beam suitable for either the Cu or Mo X-ray, respectively. The whole assembly (X-ray source, optics, and collimation system) is firmly connected and can be moved along three axes (x, y, z) to enable the X-ray beam position to be fine-tuned such that both the X-ray and neutron beam can be superimposed at the sample position. A goniometer (Huber) with six degrees of freedom to achieve highly flexible sample positioning was employed. For the experiment presented here, the sample was positioned at 45° relative to both of the orthogonal X-ray and neutron beams. The scattering data are acquired by an EIGER R 1M detector located in an evacuated detector chamber. No beamstop is needed to stop the direct beam in front of the detector, enabling scattering data and direct beam intensity measurements on the detector without any inferior scattering of the beamstop. As a result, no corrections due to the beamstop shadow is

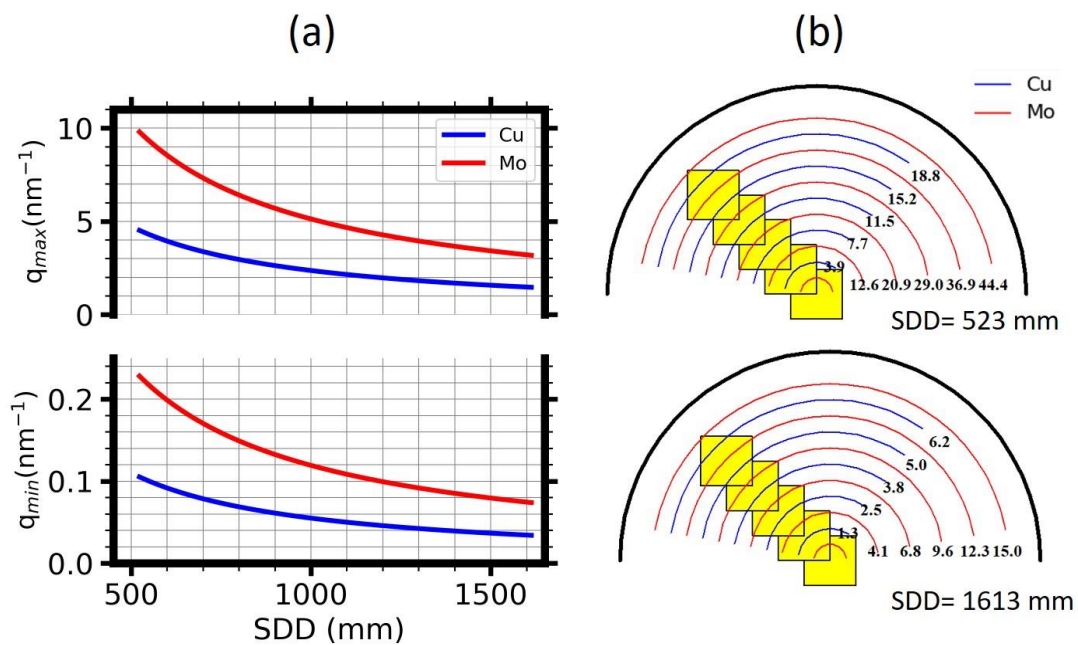

**Figure 2** a) center and b) off-center accessible q ranges at different SDDs for both Cu $K_\alpha$ and Mo $K_\alpha$ radiation. When the direct beam is on the detector's active area (center position), the $q_{min}$ (smallest scattering angle) is assumed at a value of 4*FWHM of direct beam intensity. For off-center configuration, scaled representations of the detector vacuum tube (black semicircle) and detector active area (yellow rectangle) for two different SDDs (523 mm and 1613 mm) are shown. The blue (Cu $K_\alpha$) and red (Mo $K_\alpha$) semicircles represent scattering rings at different accessible q values (nm$^{-1}$) and their relative position on the detector's active area.

required. All components of the system were mounted on a stand-alone metal rack (chassis) that made it easily movable for use on the D22 instrument. The D22 floor was equipped with an adjustable metal-concrete support base for the fast and precise positioning of the SAXS system in the D22 zone. The Network based Instrument COntrol System (NICOS) software is employed for controlling all instrument motors and data acquisition. It is a Python-based user-friendly client-server running on top of TANGO environment (Kleines *et al.*, 2015). The detector allows the X-ray direct beam either in a center or off-center position, permitting various q ranges ($q = 4\pi*\sin\theta/\lambda$, where theta is half the scattering angle). When the beam is on the active area of the detector, the range of the scattering angle for Cu K$_\alpha$ radiation of 8.048 keV (0.15406 nm) can be varied by selecting different SDDs, enabling an interactive q range between 0.040 and 4.4 nm$^{-1}$ (**Figure 2a**). While, Mo K$_\alpha$ radiation of 17.45 keV (0.0711 nm) can alternatively be used when higher penetration depth of the radiation, or a larger q (0.07-9.7 nm$^{-1}$) range, is demanded (**Figure 2a**). Various larger q ranges can be further permitted by moving the detector to an off-center position in the vacuum chamber (see **Figure 2b**).

## 3. SAXS configuration

### 3.1. X-ray source and optics

A dual wavelength (DW) X-ray source that is based on the latest generation of the microfocus rotating anode (Rigaku MicroMax-007 HFMR) generator was employed. By utilizing only one rotating anode, two switchable X-ray targets (Cu and Mo) are available in one compact system. The effective focus size is 70 μm, and the generator operates at 40 kV-30 mA for Cu ($\lambda = 0.15406$ nm) while at 50 kV-24 mA for Mo ($\lambda = 0.0711$ nm). At the click of one button, either Cu or Mo radiation can be switched via a simple mechanical sliding mechanism between metal tracks on the anode. In contrast to the Cu K$_\alpha$ radiation typically used for soft matter samples, the Mo K$_\alpha$ is chosen for highly absorbing samples as well as further extended q range. Furthermore, Mo K$_\alpha$ radiation can be used for iron-containing samples because Cu K$_\alpha$ radiation unfavorably re-emits secondary X-ray photons (fluorescence), leading to high background on the X-ray detector, which, in turn reduces the overall data quality. To deliver either monochromatic Cu K$_\alpha$ or Mo K$_\alpha$ beams, the X-ray source is coupled with an auto-switching DW optics. Rigaku's VariMax DW™ optics is composed of dual optics in a single compact housing. The optics (for Cu or Mo radiation) employ two multilayer mirrors arranged "side by side" (Montel optics) (Liu *et al.*, 2011).

### 3.2. Collimation system

A collimation length of 56 cm is composed of three compact slit systems (JJ X-ray). The first is the tungsten carbide (WC) slit system, which is suitable for both Cu K$_\alpha$ and Mo K$_\alpha$ radiation. It is placed directly after the optics. The other two motorized scatterless slit systems are integrated at the end of the fully evacuated collimation line. These scatterless slits are either Si or GaAs for Cu K$_\alpha$ and Mo K$_\alpha$ radiations, respectively. Due to the short collimation length that has been constrained by the available space in the D22 zone, a two slit collimation setup is finally employed, either W/Si or W/GaAs. The whole assembly (X-ray source, optics, and collimation system) is placed on a 3-axis table that enables a flexible adjustment of the X-ray beam position (x, y, z) to the same sample spot probed by the neutron beam.

### 3.3. Sample cell

The sample is located within an area of about 11 x 11 cm$^2$ under ambient conditions. A 6-axis goniometer was employed to adjust the sample position relative to both the X-ray and the neutron beams. A sample cell composed of two identical compartments was constructed. One compartment held the sample, whereas the other was filled with silicon oil in which a PT100 temperature sensor was inserted to monitor the cell's temperature. The sample cell system comprised a water-cooled copper holder and two attached copper blocks with an inner silver spacer that acted as a chemically resistant and thermally conductive enclosure for the sample. Mica windows (30 mm in diameter, 25 μm thick) were held in position by PEEK caps equipped with O-rings to ensure cell tightness. The windows formed a sample thickness of 1 mm in total. A sample solution could be injected via a syringe through one of the three 0.5 mm holes from the top of the cell so that the sample was sandwiched between the mica windows. The other 0.5 mm holes serve to avoid bubble formation during sample loading. Following sample injection, the three holes were properly sealed using a thin film of pressure-sensitive adhesive to minimize solution evaporation and keep the sample concentration unchanged during experiments. The sample was situated and aligned at 45° relative to both the X-ray and the neutron beam. A corrected optical path length of 1 mm*$\sqrt{2}$ for both X-rays and neutrons is calculated. The sample cell (with two compartments) could be removed from the temperature-controlled copper holder and replaced with another freshly cleaned cell to minimize downtime. The sample alignment was perfectly maintained upon sample change.

### 3.4. Detector

A dedicated in-vacuum windowless X-ray EIGER R 1M detector (Dectris) based on the hybrid pixel technology has been installed inside a large vacuum tube. The detector is composed of two modules arranged in horizontally oriented rows, yielding a total active area of 79.9 mm x 77.2 mm (array of 1030 x 1065 pixels, 75 μm pixel size). The detector position can accurately be adjusted along the three spatial directions (x, y, z). In the direction of the beam (x-direction), the detector can move up to 1.6 m distance from the sample to allow extreme SAXS ($q_{min}$ = 0.04 nm$^{-1}$) or to 0.5 m distance allowing WAXS/MAXS measurements ($q_{max}$ = 4.4 nm$^{-1}$). Furthermore, the transverse motion of the detector enables the beam to be either in a center or off-center position, permitting an extreme WAXS maximum q value of 40.0 nm$^{-1}$ (**Figure 2b**). With a continuous readout with duty cycle > 99% capabilities, EIGER R is able to acquire a short exposure time with a frame rate of 10 Hz. The transmitted beam intensity is continuously measured on the detector; no beamstop was used to stop the direct beam. With EIGER's auto-summation mode, the frames were acquired at high frame rates on the pixel level, effectively avoiding any overflows. The detector system summed the short-time-

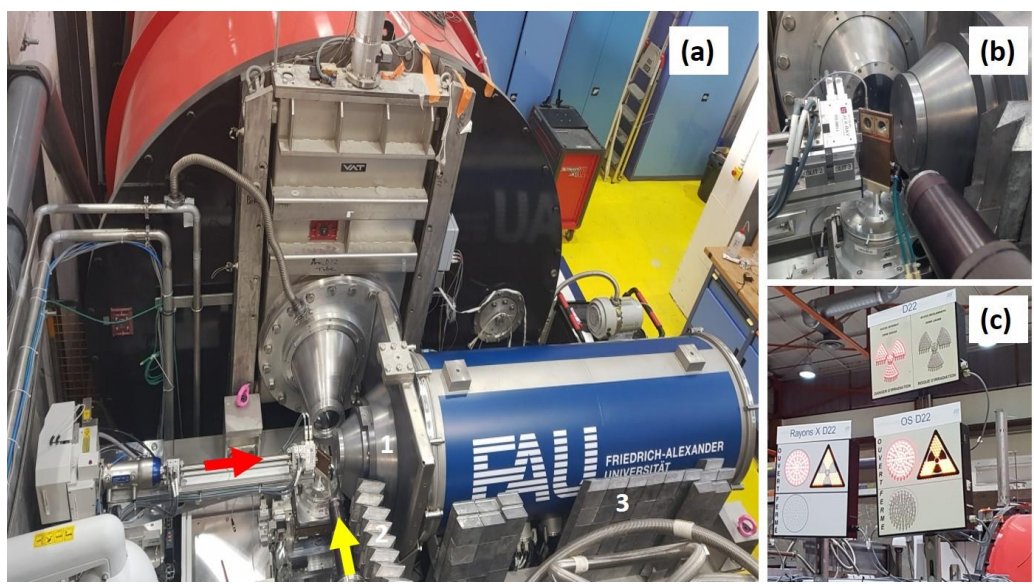

**Figure 3** a) Photograph of the SAXS instrument installed at D22 instrument (ILL) for simultaneous SAXS/SANS experiment, b) a sample at an angle of 45° relative to both orthogonal neutron and X-ray beams, c) additional zone warning lamps including safety interlocks designated for remotely operating the X-ray experiments inside the D22 zone. Lead shielding can be seen: (1) on the front of detector vacuum chamber, (2) along the neutron collimation and (3) on the side of the vacuum chamber of the X-ray detector. The red and yellow arrows indicate the directions of X-ray and neutron beams, respectively.

acquired frames (before overflow) to images on the fly, extending the bit depth of the data by the number of summed frames. This enabled direct beam intensity determination on the detector, allowing count rate up to $2*10^6$ cts/s per pixel, and prevented inconsistency of the recorded patterns due to the inevitable alternating shadowing of the beamstop.

### 3.5. Control software

Device servers for both the EIGER R 1M (Dectris) detector and phyMOTION$^R$ motor controllers were developed and implemented within the TANGO control system framework (Chaize *et al.*, 1999, Kleines *et al.*, 2015). NICOS was initially developed at FRM II by one of the coauthors and later maintained and developed by the MLZ software group. It provides a user interface for performing scientific experiments interacting with the underlying TANGO system. The NICOS user interface offers an online graphical view of the acquired data and an interactive Python command scripting for fully automated experiments.

### 3.6. Portable SAXS at D22- ILL

The transport flexibility of the current advanced laboratory-scale SAXS system opens up new opportunities for combining multiple complex and bulky analytical methods in a single experiment (**Figure 3**). Targeted to be installed at the D22 instrument (ILL) for simultaneous SAXS/SANS measurements, the SAXS system consists of a portable heavy-duty chassis that hosts all components (source, detector generator, chillers, vacuum pumps, motor controllers, main remote servers, and electrical and mechanical units). The mobility aspect of this system makes it ideal not only as a transportable system to a remote location (D22-ILL) for in-situ studies but also as a standby 24-7 independent complimentary research tool at ILL, enabling potential and important independent SAXS experiments following neutron measurements for samples that are highly susceptible to fast aging.

Additionally, to assure a quick initial adjustment of the system after the transport process inside the D22 zone, the floor of the D22 zone was equipped with two steel-concrete bases. The supporting base comprised concrete wrapping the steel base plate and a detachable vertical steel plate perpendicular to the upper surface of the base plate. With the help of adjustment screws on these perpendicular plates, the SAXS chassis could be satisfactorily adjusted at 2.5 cm range distances. Safety interlocks and necessary zone warning lamps for operating the new SAXS system at the D22 zone were already installed (**Figure 3c**). The safety interlock checks, radiation exposures, and related radiation protection challenges have all met ILL radiation safety regulations and approved by the French nuclear safety and radiation protection authority. The portable SAXS system is outfitted with a rear central hub (**Figure 1c**) that comprises 1) 16A and 32A electrical power sockets, 2) an Ethernet port, 3) sockets for a radiation safety circuit connection to the D22 zone, and 4) a port and a socket for a quick connect-release coupling of the inlet and outlet cooling water. Easy start-up of the SAXS system was thus achieved by using these plugs and sockets to enable all necessary electrical and cooling-water connections to the instruments' components. Overall, the SAXS system is mobile in design and can be used as a plug-and-play (PnP) instrument. Within one hour, the system can be craned; placed inside the D22 zone; connected to power supplies, cooling water, and the Internet; and then turned on for the alignment step. Uniquely, a conceptual change of this current custom-made SAXS system with respect to commercial systems is its high flexibility in design with all its main components (source, optics, collimation, sample, and detector) freely moveable in all directions (x, y, z). This flexibility is essential to probe the same sample volume using both X-ray and neutron radiation—placing the X-ray beam within ±1 mm of the same sample area, where neutrons are impinging on the sample. Following an experiment, the system can be easily craned out of the D22 zone, where heavy-duty castors can be attached to the SAXS chassis for lodging in a spacious booth in the experimental hall of ILL.

## 4. Performance evaluation of the portable SAXS system

A bright DW (Cu and Mo) rotating anode X-ray generator (Rigaku MicroMax-007 HFMR) and focusing optics (Rigaku's VariMax DW$^{TM}$) followed by a three-slits collimation system provide a photon flux of ~1.1x10$^7$ photons s$^{-1}$ (for 0.5 x 0.5 mm$^2$ beam on the sample position). The new generation of the Rigaku source runs at a maximum power of 1.2 kW and rotating speed of 9000 rpm and has a small effective focal point of 0.07 mm. These characteristics facilitate a performance that approaches that of second-generation synchrotron bending-magnet beamlines. Because the SAXS system is targeted to be mounted/demounted upon beamtime allocations at the D22 instrument, important characteristics such as flux, space, mobility, and quick start-up are prioritized. For instance, the beam brilliance of the currently employed microfocus rotating-anode generator source is 6 times more than that of the state-of-the-art microfocus sealed tube (Skarzynski, 2013). Although the high beam brilliance of metal-jet-based sources developed by Excillum exceed the brilliance of the current microfocus rotating anode source, the required transportable flexibility hinders the usage of these sources. The enhanced stability of the electromechanical components of the MicroMax-007 HFMR X-ray source, compared to traditional rotating anode sources (Skarzynski, 2013) operating at very high power (12 kW) and a large focal point (0.30 mm), reduces the frequency of the maintenance/inspection processes. Thus, the currently employed X-ray source provides high flux and mechanical robustness needed for the current portable lab-scale SAXS system. On the detector side, data quality, costs, and time needed for the data collection are major determining factors for the selected detector technology. The EIGER R 1M detector (Dectris) is based on hybrid pixel technology in which a "single photon" is converted into an electric charge that is processed in the complementary metal-oxide semiconductor readout chips. With no dark current or readout noise, short readout time (10 μs), excellent spatial resolution (75x75 μm$^2$), and a relatively large-sized active area (1030 x 1065 pixels), this detector is well suited for the intended combined SAXS/SANS experiments. Compared with the extremely fast readout (200 Hz) of Pilatus (Dectris) detectors suited for synchrotron radiation, a window-less and vacuum-compatible EIGER R detector with a frame rate of 10 Hz (100 μs frame time) is suited for SAXS data-collection comparable to the fast SANS measurements at D22. The counting rate of $2\times10^6$ cts/s/pixel, high frame rate, and auto-summation mode of the employed X-ray detector allow data collection (HDF file format) without the need to use a direct beamstop to attenuate the direct beam, summing all collected short-time frames for a long exposure time. Without a beamstop, no specific correction for low q data is needed. The beamstop typically limits the minimum resolution (lowest angle data) due to the associated diffused shadow on the detector. The whole system is evacuated on the optics, collimation, and detector to minimize air-scattering effects. However, at the sample position, a free space (air) is necessary to allow a spacious area for a perpendicular neutron beam (with respect to the X-ray beam).

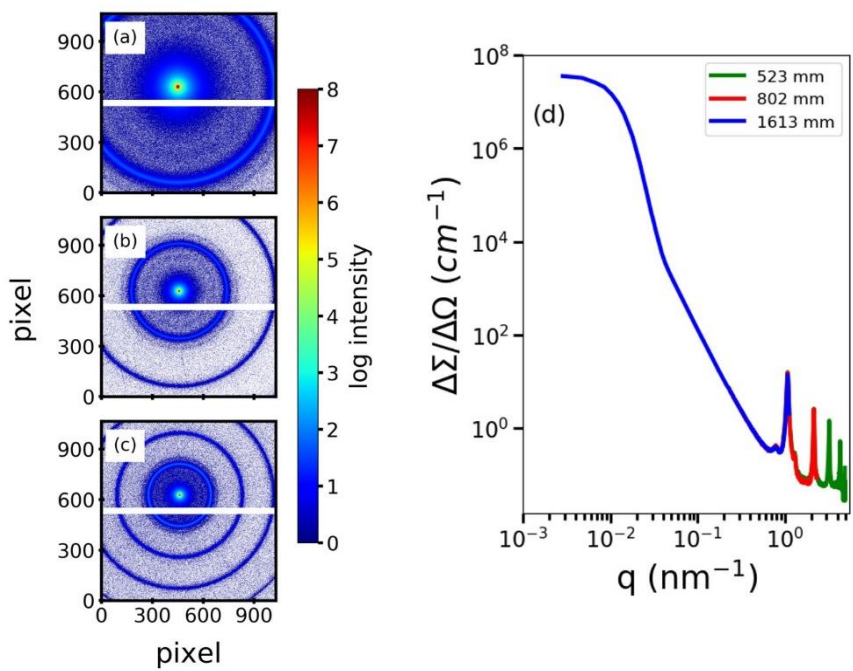

**Figure 4** 2D SAXS scattering patterns of silver behenate (Agbh) at three different SDDs of a) 1613 mm, b) 802 mm, and c) 523 mm indicating a SAXS q dynamic range between 0.04 and 4.4 nm$^{-1}$, d) Azimuthally averaged SAXS data of Agbh showing an overlapping data along the three employed SDDs.

The performance of the portable SAXS instrument was examined using calibrants and reference samples. For instance, glassy carbon, as a secondary intensity standard, and silver behenate [$CH_3(CH_2)_{20}COOAg$] have been used to calibrate the absolute intensity scale and the q-scale, respectively. The 2D SAXS scattering patterns of silver behenate at three different sample-to-detector distances (SDDs) are displayed (**Figure 4**). Having the beam in a centre position on the detector, the scattering pattern with different q ranges can be acquired for SDDs between 0.5 and 1.6 m. Due to perfect q-calibration, peaks accurately overlap for all the measurements (**Figure 4d**). Two aqueous suspension of organic and inorganic nanoparticles were used for testing our system. The first sample is an aqueous suspension of platelet-like shaped tripalmitin nanocrystals (TP NCs) solution. Corresponding systems (TP NCs) have been intensively studied by our group (Schmiele *et al.*, 2015, Schmiele, Gehrer*, et al.*, 2014) as potential drug delivery systems. The second sample is an aqueous dispersion of commercial silica particles (Ludox TM50, 50 wt%) purchased from Sigma-Aldrich, the average diameter of the $SiO_2$ NPs is about 26±2 nm (see fitting results in **Figure 1S**). Both samples are selected because they provide two different X-ray scattering power (organic and inorganic particles) and covering essentially different q ranges (**Figure 5**).

The dispersions were filled in the sample cell that is a 1-mm thickness gap sealed on both sides by two thin parallel mica windows. The sample holder was installed at the SAXS system in an air gap of about 11 cm. It was also apparent that the corrections for background and detector noise have minimal effect on the data. With a beam size of 0.5 mm x 0.5 mm on the sample position and a divergence of about 0.4 mrad at the detector position (SDD 1612 mm), the apparent minimum value of q for which usable intensity can be obtained is 0.04 nm$^{-1}$. Such excellent low $q_{min}$ despite the short collimation is achievably due to absence of a beamstop. when the beam is in a center position on the active area of the detector, two SDD distances

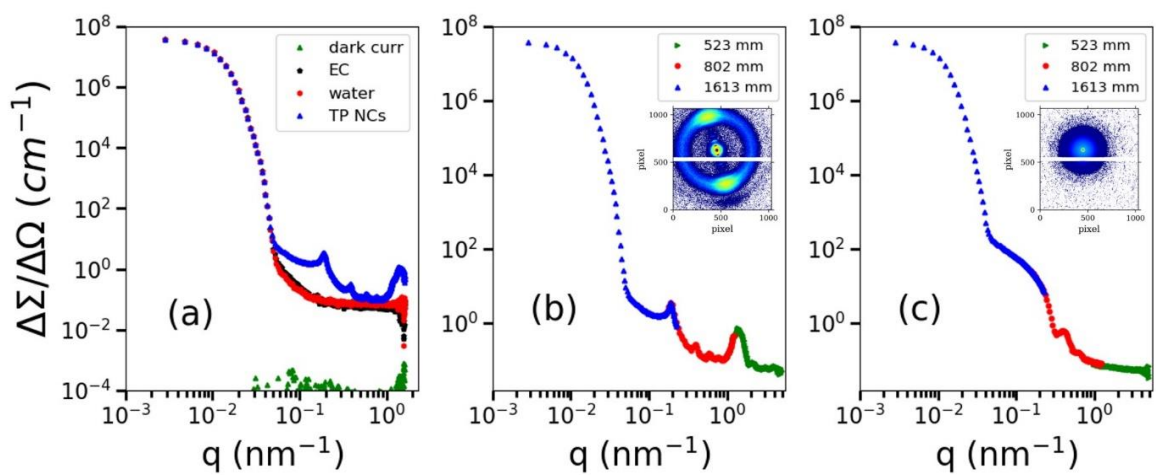

**Figure 5** a) Azimuthally averaged SAXS data of empty cell EC (black), a 1-mm thick water (red) sandwiched between two mica windows (cell), the detector dark noise (green), and a sample of platelet-like shaped tripalmitin nanocrystals (TP NCs) in water (blue) collected at SDD 1613 mm. The 1D SAXS profiles of b) platelet-like shaped tripalmitin nanocrystals (TP NCs), c) colloidal silica NPs (Ludox TM50; average size = 26±2 nm) collected at different SDDs. Data were plotted and overlapped over different q range on an absolute scale.

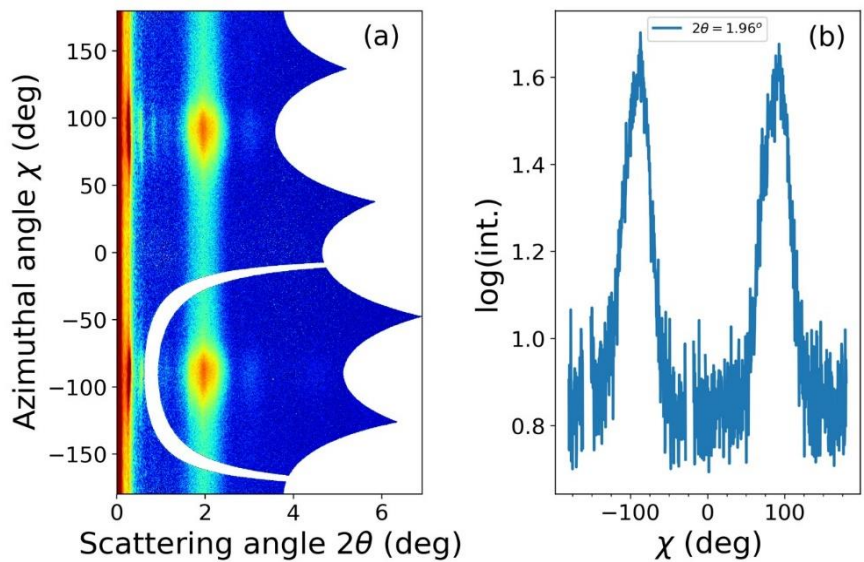

**Figure 6** a) Azimuthal 2D regrouping, b) radial plot (at $2\theta = 1.96°$) of platelet-like shaped tripalmitin nanocrystals (TP NCs) collected at a SDD of 523 mm.

give access to a q range between 0.04 and 4.4 nm$^{-1}$, (**Figure 2**a) while 2θ values are in the range between 0.2 and 6.5° (**Figure 6**). An off-center detector configuration (direct beam is not on the active area of the detector) can further be used for a wide angle scattering experiment, permitting a maximum q of about 18.5 and 40.0 nm$^{-1}$ (**Figure 2b**) for Cu K$_\alpha$ and Mo K$_\alpha$ radiation, respectively.

## 5. SAXS/SANS setup

The SAXS system has fulfilled all requirements including geometrical constraints, transport flexibility, and safety aspects for installation at the D22 instrument. However, to ensure feasible and excellent operation, further challenges need to be overcome: 1) managing high energy gamma and X-ray radiation levels at the ILL instrumental zones to acquire low background 2D scattering patterns on the X-ray detector, 2) validating the SAXS and SANS data for the samples collected at a 45° angle to those collected at 90° (typical sample angle for conventional scattering experiments), 3) enabling a unified control of both SAXS and SANS experiments using user-friendly ILL software tools as well as developing an appropriate SAXS/SANS data reduction/analysis software using unified data set formats and fitting models. In the current work, we emphasize only the technical aspects of the SAXS system and the realization of combined SAXS/SANS measurements in one experiment. The software development of the state-of-the-art simultaneous SAXS/SANS method is in progress and not the scope of the current work.

### 5.1. SAXS and background radiation at D22

We explored the spatial distribution of high energy gamma radiation in the D22 zone and constructed a preliminary experimental lead shielding around the SAXS system. This includes radiation protection walls with a thickness of 7-10 cm at the zone walls (between the D22 zone and other experimental zones), as well as lead shielding along the neutron collimation (**Figure 3**). On the front side of the evacuated detector chamber of the SAXS setup, we designed, constructed, and installed a cone of lead-based (PbSb$_4$) material. In addition, a lead wall right on the SAXS chassis facing the neutron guide was installed (**Figure 3**). Preliminary test experiments on the SAXS detector at D22, with the neutron source shutter either open or close, were extensively performed at different neutron collimation lengths and X-ray SDDs. Following the implementation of the lead shielding, a significant reduction of the inferior background radiation on SAXS detector was achieved. In particular, a radiation background level competitive to the SAXS setup outside a neutron facility could already be achieved for long neutron collimation lengths. **Figure 7** compares the 1D SAXS scattering profiles of monodisperse 100-nm silica NPs for both opened (red) and closed (blue) neutron shutters at different neutron collimation lengths and X-ray SDDs. The highest background radiation has been observed for a configuration with a short neutron collimation length (2.8 m) and a small X-ray SD distance (0.5 m) (**Figure 7c**).

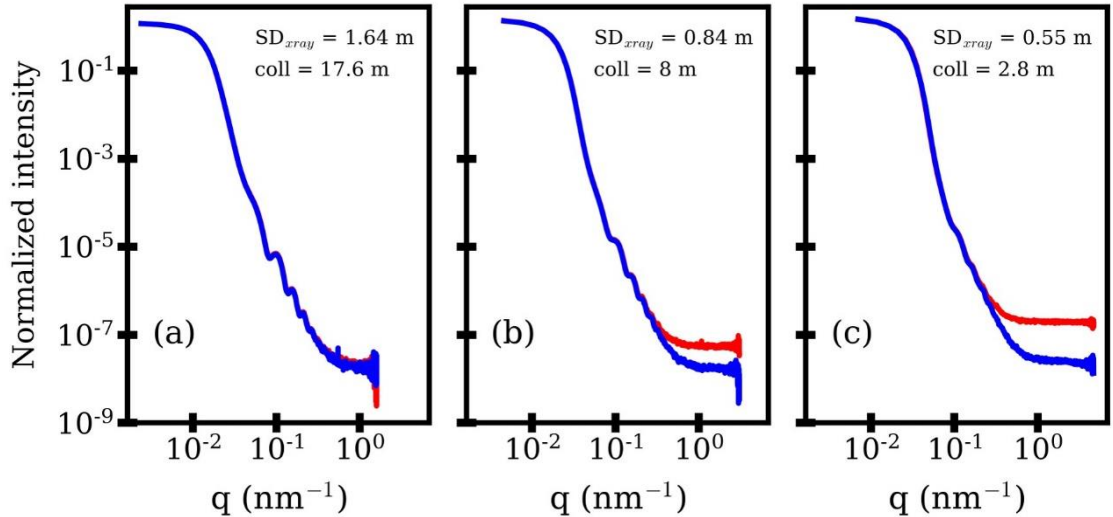

**Figure 7** 1D SAXS profiles calculated from the 2D detector images of silica NPs (100 nm) recorded with an opened (red; simultaneous SAXS/SANS mode) and a closed (blue; standalone SAXS mode) neutron shutter for different neutron collimation lengths and X-ray sample-to-detector ($SD_{x-ray}$) distances. Due to the reduced neutron flux, the radiation background is insignificant on the X-ray detector for a) long collimation length, while for b) intermediate and c) short collimation lengths a significant background radiation is observed for $q > 0.31$ nm$^{-1}$. The intensity is plotted as an intensity normalized to the direct beam intensity with no background radiation subtraction for comparison purposes.

The current results indicate, however, the need for further reduction in background radiation (about 3-10 times) for intermediate and short neutron collimation lengths. Very recently, a double-threshold energy detector has been developed by Dectris (EIGER2 X 1M) and is indeed an optional upgrade to further improvements (see Section 7). The preliminary results on this detector (data are not shown) operated at the D22 zone have confirmed an efficient background-radiation reduction of about 10-15 times by employing, for instance, an energy window (7-9 keV) is better suited for the employed option of the Rigaku X-ray source at 8 keV X-rays. As part of the development process, such a detector will be purchased and installed. Together with further studies related to the optimum shielding configurations, state-of-the-art laboratory SAXS data quality for all currently employed SANS setups at D22 for all operation conditions (collimation length, aperture, and various beam flux) should be achievable without a special background subtraction.

### 5.2. SAS at two different sample angles

For the SAS method, the beam is typically on the sample at a normal incidence. For the combined SAXS/SANS experiment, both X-ray and neutron beams need to simultaneously illuminate the same sample volume. Thus, the sample should be positioned at an angle that allows dual-beam-based SAS investigations. One possible simple geometry is to place a cuvette at a 45° angle relative to both orthogonal X-ray and neutron beams. Our results indicate that a simple geometrical correction (sample thickness*√2) can be used to replicate the X-ray (**Figure S2**) and neutron (**Figure 8**) scattering data of a conventional normal incidence beam experiment.

## 5.3. In-situ simultaneous SAXS/SANS experiment

The first science experiment using simultaneous SAXS/SANS technique has been successfully performed at D22-ILL. The cationic micelle structure development and evolved Au NPs growth behavior during Au NPs synthesis, via seed-mediated growth procedure, were simultaneously monitored to emphasize the capabilities of the truly combined SAXS and SANS methods. Related preparation protocols for preparation of rod-shaped Au NPs have previously been published by our group (Schmutzler, Schindler, Schmiele, et al., 2018, Schmutzler et al., 2019). The Au NPs were synthesized using gold precursor ($HAuCl_4$) that forms small seeds after an initial reduction step using reducing agent $NaBH_4$ in the presence of cetyltrimethylammonium bromide (CTAB). The subsequent reduction of $HAuCl_4$ using hydroquinone ($C_6H_6(OH)_2$) in the presence of $AgNO_3$, CTAB, and the seed particles resulted in the formation of single crystalline rod-shaped Au NPs. Within 2 hours, the stabilized large Au NPs were mostly obtained during the reduction step at a marginally elevated temperature (~ 35°). Structural determination of the prepared sample was examined using simultaneous SAXS/SANS methods probing the same sample volume. This indeed would provide new insights onto possible structural reorganization of the stabilizing CTAB agent in concomitant with the gold particle growth behavior. The developed simultaneous SAXS/SANS method is perfectly suited to identify such correlation among individual-specific behaviors (organic micelle and inorganic particle). Using simultaneous SAXS/SANS method, the evolved particle growth (SAXS) and the CTAB structural modifications (SANS) were monitored at a time resolution of one minute. The time evolution of both 2D SAXS and SANS patterns is visualized in the insets of **Figure 9a,b**. A gold characteristic SAXS signal begins to be distinguished at about 55 min of the synthesis process (**Figure 9a**). As revealed from the SANS data, a characteristic structural peak of the CTAB micelles (**Figure 9b**) at q of 0.46 nm$^{-1}$ was slightly shifted to higher value of 0.50 nm$^{-1}$ in a few minutes. Followings, an observed fast decay of the SANS signal was significantly deaccelerated when the first signs of the characteristic gold SAXS signal was detected within 55-60 min. After that, the SANS signal was diminished only slightly and leveled out at around 120 min. Interestingly, this coincides with a cease in the SAXS intensity change (at 120 min). The SANS and SAXS results indicated a significant reduced volume fraction of micelles (**Figure 9b**) in concomitant with a progressive formation of a defined Au NPs shape and size within the first 55-60 min (**Figure 9a**). All details of sample preparations, the employed model, and data fitting procedure are described in the supporting information. The SANS signal reached 1/e of its initial intensity after about 55 min, as it did not show any significant structural modifications during particle formation. In contrast, the SAXS signal intensity continued to develop up to 120 min into the synthesis. The Au particle structure extracted from the SAXS data can be described as mostly rod-shaped or elongated particles with an average width and length of about 8 nm and 60 nm, respectively.

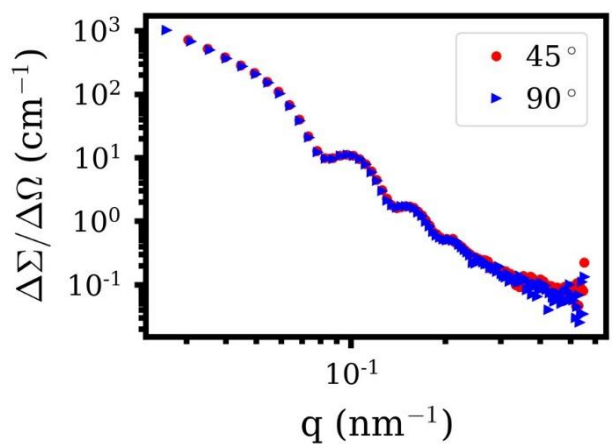

**Figure 8** 1D SANS profiles of 100-nm silica NPs collected at two different sample angles. Correction of scattering profile, by multiplying each data point of the sample measured at 45° angle by 1 mm* $\sqrt{2}$ , is perfectly coincidence with that collected at 90° angle.

A prominent large Au particle size development can be particularly assumed during the first 55-60 min. The latter stages of Au NP growth above 60 min can be interpreted as a spontaneous restructuring phase of the evolved gold nanoparticles, where gold atoms and seeds are rapidly colliding together forming large sized particles as indicated from the enhanced SAXS signal intensity after 55 min. For the Au NP, the initial growth within the first 55 min could not be detected due to the low scattering volume. It seems, the incorporation of CTAB micelles in a structurally different stabilization shell is happening in the early stages of the synthesis. During the formation of large Au NPs, the CTAB micelle characteristic SANS signal dramatically diminishes. Then it is slightly changed, while the volume fraction of structurally well-defined Au particles continuously increases (starting at 55-60 min) via a collision mechanism (Schmutzler *et al.*, 2019).

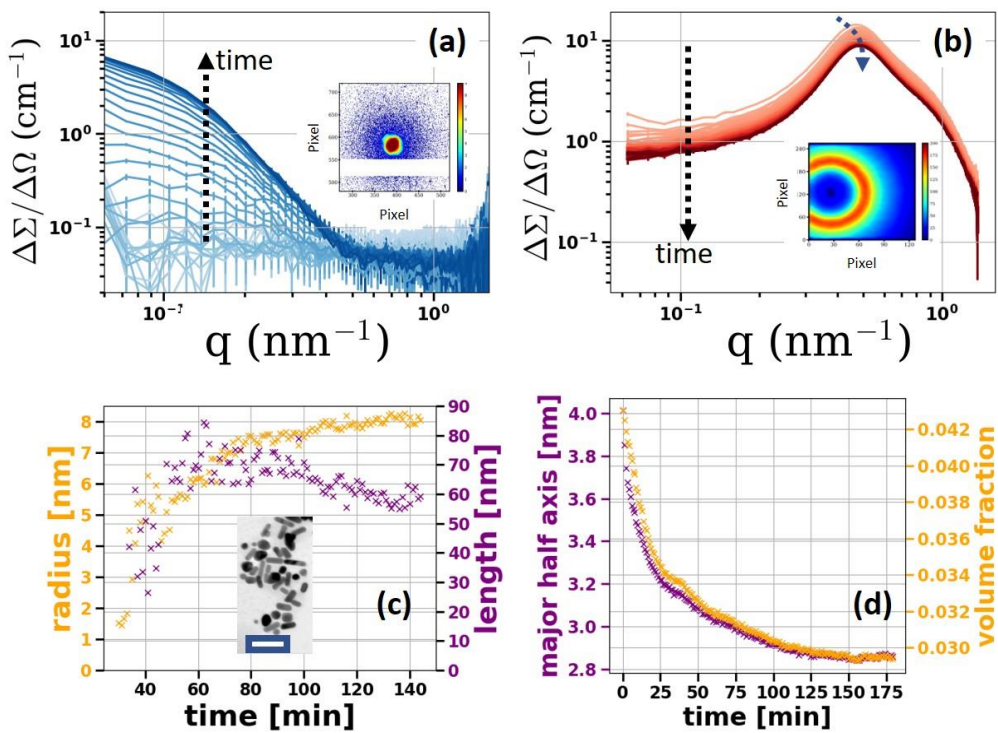

**Figure 9** Time evolution of both a) SAXS and b) SANS 1D profiles during the in-situ reduction process of a gold solution at 35°C. Insets in a) SAXS and b) SANS 2D scattering patterns during hydroquinone synthesis of Au NPs. c) Rod-like Au particles' radius and length as a function of reaction time, as revealed from the SAXS data fittings. The insert in c) TEM image of Au particles with average radius and length of about 8 nm and 60 nm, respectively; the presented scale bar is 100 nm. d) An exponential volume fraction decay of the CTAB micelle with time, as obtained from SANS data fitting.

Therefore, it is apparent that potential structural reorganization of the stabilizing agents during the NPs formation is not a major player in determining the size/shape of the evolving Au NPs. Our previous studies have already confirmed, that the micelle structure influences the size and aspect ratio of gold NPs and can influence the stabilization of Au NP in solution tremendously depending on their morphology (Schmutzler *et al.*, 2019). Consistently, the observations made in this measurement mean that even though the Au NPs grew drastically (55-60 min), the micellar structure, residing at the surface of Au NPs or free in solution, is maintained. This clearly rules out particle stabilization mechanisms solely via formation of bilayer on the particle surface, which has been mainly discussed in literatures for a long time. Indeed, the stability of the intermediate Au NPs is established via an attachment of the CTAB micelles themselves from solution to the surface of the Au particles. The so attached micelles might regulate the Au NP growth relative to certain crystal facets via steric stabilization and manipulate the reaction and the subsequent attachment of gold ions to the surface, therefore driving an anisotropic metal growth.

Now, we are able to monitor the size/shape evolution of the CTAB micelles, Au NP growth behavior, and how they are linked to each other, further simultaneous SAXS/SANS measurements can be envisioned. So far, many inconsistent and controversial structural models from studies on the correlation between size/shape of the structures formed by the organic stabilizer (e.g. CTAB) and the evolved inorganic NPs (e.g. Au) morphology have been reported (Karayil *et al.*, 2015, Ye *et al.*, 2012). This is due in part to the poorly studied interplay between organic stabilizer and Au NPs during the synthesis step. The current results indeed support our recent theory regarding the role of the micelle structure on the evolved particle morphology. Furthermore, it is feasible to further investigate the evolution of Au NPs morphology corresponding to the size/shape of CTAB micelles in presence of additives (n-alcohol; hexanol, pentanol or decanol). The addition of n-alcohols increases the micelle size/number and therefore is presumed to potentially alter the facet dependent growth of Au NP, which in turn influences size and shape without changing the overall chemistry of the synthesis. This will be an interesting approach for further exploration on these systems and other relevant functional systems using our advanced simultaneous SAXS/SANS method.

## 6. Conclusions

Combining SAXS and SANS methods not only prompts a synergy that results when data sets are obtained simultaneously from the same sample volume but also allows access to the cross-correlation of phases/components when two different simultaneous contrast situations are probed for multi-phase/component samples, ideally during in-situ/real-time experiments. Moreover, the SAXS/SANS system will enable contrast-dependent successive length-scale characterization of novel nanomaterials, addressing unsolved scientific questions. This study presented the technical characteristics of a newly established portable SAXS apparatus that is dimensionally suitable for implementation at the D22 instrument of ILL, enabling combined SAXS and SANS measurements. The SAXS instrument was based on a high flux rotating anode generator with selectable Mo/Cu $K_\alpha$, enabling nanoscale structural studies of soft materials and biological systems. The SAXS system comprised a collimation system including a Mortel multilayer optics and scatterless slits, sample stage with six motional degrees of freedom, evacuated beam path combined with an ambient sample holder, and EIGER R 1M detector. The source-sample-detector geometry resulted in scalability and flexibility to expedite the alignment step of the X-ray beam on a distinct sample volume, which is probed by neutrons. A heavy-duty non-magnetic metal chassis was employed to incorporate all the system's components. The SAXS apparatus, which is mobile and peripheral in design to support PnP operations, enables its quick (< 1 h) installation at the D22 instrument. Software control based on the NICOS user interface on top of a TANGO interface is used to control all motors and collect data in a fully automated manner. The assessment of the SAXS performance has been presented here for the Cu $K_\alpha$ radiation. Alternatively, the Mo $K_\alpha$ radiation can be used as needed for alternative sample geometries. A wide q dynamic range can be achieved via a movable detector inside the vacuum chamber. Background intensity on the SAXS detector mainly caused by hard gamma rays could be efficiently minimized by effective optimized lead shielding. This study presented a test experiment to emphasize the unique capability of in-situ SAXS/SANS studies for determination of structural rearrangements during materials processing such as controlled nanoparticle growth. Exemplarily, the structural determination of both the hard (Au NPs) and soft moieties (CTAB) of surfactant-stabilized Au particles during the Au NPs growth process was performed. The temporal evolution of the particle shape, size, and their stabilization layer and structural rearrangements of the CTAB micelles can only be determined and temporally correlated by using simultaneous SAXS and SANS methods.

## 7. Future developments

Optimization for fully automated and simple operating simultaneous SAXS/SANS measurements is a requirement to allow routine operations at D22 (ILL). The ultimate aim is to offer a simultaneous SAXS/SANS method as a widely used standard method for soft matter research and beyond at ILL. Thus, further developments, including sample environments, multi-sample changer, optimized routine operation, and user-friendly software, should be considered in future research. Few essential sample environments (Jordan *et al.*, 2016, Lopez *et al.*, 2018, Wang *et al.*, 2004, Weigandt *et al.*, 2011) that are currently employed at ILL should be modified/developed to fit our system and to satisfy the demand of the ILL user community. However, a multiple-sample changer with temperature control should be constructed for routine sample measurements. Further, the sample environments such as HPLC and flow cell already available at ILL are proposed to be modified to fit the combined setup to simultaneously collect X-ray and neutron data on biological samples, such as protein samples prone to aggregation. As already mentioned above, a detector with a double-energy threshold will be purchased to enable minimum background radiation on the X-ray detector. Moreover, standard data acquisition/monitoring routines, data format, and relevant data analysis software have to be developed. The optimization process also includes the standardization of the SAXS operation for a flexible monitoring/data acquisition within the ILL software tools. In particular, a common flexible control software interface with several predefined standard configurations to enable small, medium, or wide angle X-ray scattering operation synchronized with the SANS data acquisition needs to be developed.

**Acknowledgments:**

We acknowledge the financial support by the Federal Ministry of Education and Research of Germany (BMBF) in the framework of project number 05K16WE1. We also gratefully acknowledge the funding of the Deutsche Forschungsgemeinschaft (DFG) through the "Cluster of Excellence Engineering of Advanced Materials (EAM)", the research training group GRK 1896 "In-Situ Microscopy with Electrons, X-rays and Scanning Probes" and the research unit FOR 1878 "Functional Molecular Structures on Complex Oxide Surfaces". We thank Dr. Georg Brandl (MLZ, TU München) for his continuous support with the integration of NICOS and TANGO onto our system.

**References**

Arleth, L., Jensen, G. V., Huda, P., Skou, S., Shang, W. F. & Chakravarthy, S. (2014). *Acta Crystallogr A* **70**, C609-C609.
Bergstrom, M. & Pedersen, J. S. (2000). *J Phys Chem B* **104**, 4155-4163.
Blanchet, C. E. & Svergun, D. I. (2013). *Annu Rev Phys Chem* **64**, 37-54.
Chaize, J. M., A.G., G., Klotz, W. D., Meyer, J., Perez, M. & Taurel, E. (1999). International Conference on Accelerator and Large Experimental Physics Control Systems, 474-479.
Chen, Y. J., Tokuda, J. M., Topping, T., Sutton, J. L., Meisburger, S. P., Pabit, S. A., Gloss, L. M. & Pollack, L. (2014). *Nucleic Acids Res* **42**, 8767-8776.
Fitter, J., Gutberlet, T., Katsaras, J. & Fitter, J. (2006). *Neutron scattering in biology : techniques and applications.* Berlin ; New York: Springer.
Frank, J. (2002). *Annu Rev Bioph Biom* **31**, 303-319.
Futscher, M. H., Schultz, T., Frisch, J., Ralaiarisoa, M., Metwalli, E., Nardi, M. V., Müller-Buschbaum, P. & Koch, N. (2019). *J Phys-Condens Mat* **31**.
Gehrer, S., Schmiele, M., Westermann, M., Steiniger, F. & Unruh, T. (2014). *J Phys Chem B* **118**, 11387-11396.
Glatter, O. & Kratky, O. (1982). *Small angle x-ray scattering.* London ; New York: Academic Press.
Graceffa, R., Nobrega, R. P., Barrea, R. A., Kathuria, S. V., Chakravarthy, S., Bilsel, O. & Irving, T. C. (2013). *J Synchrotron Radiat* **20**, 820-825.
Grishaev, A., Anthis, N. J. & Clore, G. M. (2012). *J Am Chem Soc* **134**, 14686-14689.
Grossmann, J. G., Callaghan, A. J., Marcaida, M. J., Luisi, B. F., Alcock, F. H. & Tokatlidis, K. (2008). *Eur Biophys J Biophy* **37**, 603-611.
Hennig, J., Wang, I., Sonntag, M., Gabel, F. & Sattler, M. (2013). *J Biomol Nmr* **56**, 17-30.
Illing, A. & Unruh, T. (2004). *Int J Pharmaceut* **284**, 123-131.
Jensen, M. H., Toft, K. N., David, G., Havelund, S., Perez, J. & Vestergaard, B. (2010). *J Synchrotron Radiat* **17**, 769-773.
Jordan, A., Jacques, M., Merrick, C., Devos, J., Forsyth, V. T., Porcar, L. & Martel, A. (2016). *J Appl Crystallogr* **49**, 2015-2020.

Karayil, J., Kumar, S., Hassan, P. A., Talmon, Y. & Sreejith, L. (2015). *Rsc Adv* **5**, 12434-12441.
Kleines, H., Drochner, M., Wagener, M., Fleischhauer-Fuss, L., Keuler, S., Suxdorf, F., Möller, R., Janasche, S., van Waasen, S., Mertens, K.-H., Bednarek, M., Bussmann, K. & Su, Y. (2015). Proceedings of ICALEPCS, 236-239.
Koch, M. H. J., Svergun, D. I., Gabriel, A., Goderis, B. & Unruh, T. (2001). *Abstr Pap Am Chem S* **222**, U355-U355.
Koch, M. H. J., Vachette, P. & Svergun, D. I. (2003). *Q Rev Biophys* **36**, 147-227.
Kuwamoto, S., Akiyama, S. & Fujisawa, T. (2004). *J Synchrotron Radiat* **11**, 462-468.
Lapinaite, A., Simon, B., Skjaerven, L., Rakwalska-Bange, M., Gabel, F. & Carlomagno, T. (2013). *Nature* **502**, 519-+.
Liu, W. J., Ice, G. E., Assoufid, L., Liu, C. A., Shi, B., Khachatryan, R., Qian, J., Zschack, P., Tischler, J. Z. & Choi, J. Y. (2011). *J Synchrotron Radiat* **18**, 575-579.
Lopez, C. G., Watanabe, T., Adamo, M., Martel, A., Porcar, L. & Cabral, J. T. (2018). *J Appl Crystallogr* **51**, 570-583.
Metwalli, E., Rasool, M., Brunner, S. & Müller-Buschbaum, P. (2015). *Chemphyschem* **16**, 2882-2889.
Mohl, G. E., Metwalli, E. & Müller-Buschbaum, P. (2018). *Acs Energy Lett* **3**, 1525-1530.
Neylon, C. (2008). *Eur Biophys J Biophy* **37**, 531-541.
Putnam, C. D., Hammel, M., Hura, G. L. & Tainer, J. A. (2007). *Q Rev Biophys* **40**, 191-285.
Schindler, T., Gonzalez, A., Boopathi, R., Roda, M. M., Romero-Santacreu, L., Wildes, A., Porcar, L., Martel, A., Theodorakopoulos, N., Cuesta-Lopez, S., Angelov, D., Unruh, T. & Peyrard, M. (2018). *Phys Rev E* **98**.
Schindler, T., Lin, W., Schmutzler, T., Lindner, P., Peukert, W., Segets, D. & Unruh, T. (2019). *Chemnanomat* **5**, 116-123.
Schindler, T., Schmiele, M., Schmutzler, T., Kassar, T., Segets, D., Peukert, W., Radulescu, A., Kriele, A., Gilles, R. & Unruh, T. (2015). *Langmuir* **31**, 10130-10136.
Schindler, T., Schmutzler, T., Schmiele, M., Lin, W., Segets, D., Peukert, W., Appavou, M. S., Kriele, A., Gilles, R. & Unruh, T. (2017). *J Colloid Interf Sci* **504**, 356-362.
Schmidt, P. W. (1995). *Modern Aspects of Small-Angle Scattering.,* edited by B. H., p. p. 1: Kluwer Academic Publishers.
Schmiele, M., Busch, S., Morhenn, H., Schindler, T., Schmutzler, T., Schweins, R., Lindner, P., Boesecke, P., Westermann, M., Steiniger, F., Funari, S. S. & Unruh, T. (2016). *J Phys Chem B* **120**, 5505-5512.
Schmiele, M., Gehrer, S., Westermann, M., Steiniger, F. & Unruh, T. (2014). *J Chem Phys* **140**.
Schmiele, M., Knittel, C., Unruh, T., Busch, S., Morhenn, H., Boesecke, P., Funari, S. S., Schweins, R., Lindner, P., Westermann, M. & Steiniger, F. (2015). *Phys Chem Chem Phys* **17**, 17939-17956.
Schmiele, M., Schindler, T., Westermann, M., Steiniger, F., Radulescu, A., Kriele, A., Gilles, R. & Unruh, T. (2014). *J Phys Chem B* **118**, 8808-8818.
Schmutzler, T., Schindler, T., Goetz, K., Appavou, M. S., Lindner, P., Prevost, S. & Unruh, T. (2018). *J Phys-Condens Mat* **30**.
Schmutzler, T., Schindler, T., Schmiele, M., Appavou, M. S., Lages, S., Kriele, A., Gilles, R. & Unruh, T. (2018). *Colloid Surface A* **543**, 56-63.
Schmutzler, T., Schindler, T., Zech, T., Lages, S., Thoma, M., Appavou, M.-S., Peukert, W., Spiecker, E. & Unruh, T. (2019). *ACS Applied Nano Materials* **2**, 3206-3219.
Schonhals, A., Frunza, S., Frunza, L., Unruh, T., Frick, B. & Zorn, R. (2010). *Eur Phys J-Spec Top* **189**, 251-255.
Schuldes, I., Noll, D. M., Schindler, T., Zech, T., Götz, K., Appavou, M.-S., Boesecke, P., Steiniger, F., Schulz, P. S. & Unruh, T. (2019). *Langmuir* **35**, 13578-13587.
Skarzynski, T. (2013). *Acta Crystallogr D* **69**, 1283-1288.
Spinozzi, F., Ceccone, G., Moretti, P., Campanella, G., Ferrero, C., Combet, S., Ojeajimenez, I. & Ghigna, P. (2017). *Langmuir* **33**, 2248-2256.

Sugiyama, M., Yagi, H., Ishii, K., Porcar, L., Martel, A., Oyama, K., Noda, M., Yunoki, Y., Murakami, R., Inoue, R., Sato, N., Oba, Y., Terauchi, K., Uchiyama, S. & Kato, K. (2016). *Sci Rep-Uk* **6**.
Svergun, D. I. (2010). *Biol Chem* **391**, 737-743.
Svergun, D. I., Feĭgin, L. A. & Taylor, G. W. (1987). *Structure analysis by small-angle x-ray and neutron scattering*. New York: Plenum Press.
Svergun, D. I., Koch, M. H. J., Timmins, P. A. & May, R. P. (2013). *Small angle x-ray and neutron scattering from solutions of biological macromolecules*, First Edition. ed. Oxford: Oxford University Press.
Svergun, D. I. & Nierhaus, K. H. (2000). *J Biol Chem* **275**, 14432-14439.
Tanaka, H., Koizumi, S., Hashimoto, T., Itoh, H., Satoh, M., Naka, K. & Chujo, Y. (2007). *Macromolecules* **40**, 4327-4337.
Unruh, T. (2007). *J Appl Crystallogr* **40**, 1008-1018.
Wang, W., Gu, B. H., Liang, L. Y., Hamilton, W. A., Butler, P. D. & Porcar, L. (2004). *Abstr Pap Am Chem S* **227**, U1208-U1209.
Wang, X. Y., Meng, J. Q., Wang, M. M., Xiao, Y., Liu, R., Xia, Y. G., Yao, Y., Metwalli, E., Zhang, Q., Qiu, B., Liu, Z. P., Pan, J., Sun, L. D., Yan, C. H., Müller-Buschbaum, P. & Cheng, Y. J. (2015). *Acs Appl Mater Inter* **7**, 24247-24255.
Wang, X. Y., Zhao, D., Wang, C., Xia, Y. G., Jiang, W. S., Xia, S. L., Yin, S. S., Zuo, X. X., Metwalli, E., Xiao, Y., Sun, Z. C., Zhu, J., Müller-Buschbaum, P. & Cheng, Y. J. (2019). *Chem-Asian J* **14**, 2169-2169.
Weigandt, K. M., Porcar, L. & Pozzo, D. C. (2011). *Soft Matter* **7**, 9992-10000.
Whitten, A. E., Cai, S. Z. & Trewhella, J. (2008). *J Appl Crystallogr* **41**, 222-226.
Whitten, A. E., Jeffries, C. M., Harris, S. P. & Trewhella, J. (2008). *Acta Crystallogr A* **64**, C297-C297.
Ye, X. C., Jin, L. H., Caglayan, H., Chen, J., Xing, G. Z., Zheng, C., Doan-Nguyen, V., Kang, Y. J., Engheta, N., Kagan, C. R. & Murray, C. B. (2012). *Acs Nano* **6**, 2804-2817.
Yin, S. S., Zhao, D., Ji, Q., Xia, Y. G., Xia, S. L., Wang, X. M., Wang, M. M., Ban, J. Z., Zhang, Y., Metwalli, E., Wang, X. Y., Xiao, Y., Zuo, X. X., Xie, S., Fang, K., Liang, S. Z., Zheng, L. Y., Qiu, B., Yang, Z. H., Lin, Y. C., Chen, L., Wang, C. D., Liu, Z. P., Zhu, J., Müller-Buschbaum, P. & Cheng, Y. J. (2018). *Acs Nano* **12**, 861-875.
Zemb, T. & Diat, O. (2010). *Journal of Physics: Conference Series* **247**, 012002.
Zheng, L. Y., Wang, X. Y., Xia, Y. G., Xia, S. L., Metwalli, E., Qiu, B., Ji, Q., Yin, S. S., Xie, S., Fang, K., Liang, S. Z., Wang, M. M., Zuo, X. X., Xiao, Y., Liu, Z. P., Zhu, J., Müller-Buschbaum, P. & Cheng, Y. J. (2018). *Acs Appl Mater Inter* **10**, 2591-2602.

## 8. Experimental

### 8.1. Materials:

Hydrogen tetrachloraurate trihydrate (HAuCl$_4$·3H$_2$O, 99.99%, Alfa Aesar), hexadecyltrimethylammonium bromide (CTAB, for molecular biology, ≥ 99%, Sigma-Aldrich), silver nitrate (AgNO$_3$, ACS, 99.9 %, Alfa Aesar), n-hexanol (≥ 99%, Sigma-Aldrich), hydroquinone (HQ, ≥ 99%, Sigma-Aldrich) and sodium borohydride (NaBH$_4$, ≥ 98%, Merck KGaA) were used without further purification.

### 8.2. Particle Synthesis:

All solutions were prepared in deuterium oxide (D$_2$O, ≥ 99.9%, Eurisotop). To synthesize the initial seed particles, a 7.5 mL aqueous, 0.1 mol/L CTAB, and 0.1 mol/L hexanol solutions were prepared and kept at 35 °C. A volume of 0.2 mL of 0.010 mol/L HAuCl$_4$ was added, which resulted a light orange solution which stems from the complexation of gold with bromide ions. Under vigorous stirring 0.6 mL of freshly prepared 0.010 mol/L NaBH$_4$ was injected changing the liquid into a brown color. The solution was left undisturbed for 30 minutes to give enough time for the NaBH$_4$ to completely decompose. Due to presence of hexanol in the seed solution, the aggregation of gold seed particles can be slowed down immensely. This way the seed solution can be used for several weeks as has previously been demonstrated (Schmutzler *et al.*, 2019). This way, the resulting Au NPs solutions can be compared with one another without any noticeable aggregation of the seed particles, enabling a much larger timeframe in which the experiments can be carried out. The in-situ experiments were performed in the custom-made SAXS/SANS sample holder with two compartments and thermally controlled copper block. Volumes of 25 µL of 0.1 mol/L HAuCl$_4$ and 25 µL of 0.1 mol/L AgNO$_3$ were added to a 7.2 mL solution containing 0.1 mol/L CTAB turning it orange. Hydroquinone was chosen as a reducing agent, because of the slower reaction kinetics as has been shown in a previous study (Vigderman & Zubarev, 2013). Following this step, a 250 µL of 0.1 mol/L HQ solution was added to the mixed solution and shacked until it turned colorless. Following the addition of 1 mL of seed solution, the slow growth of AuNP begins. The sample cell was quickly filled via syringe and inserted into the sample holder. The temperature during the in-situ SAXS/SANS measurement was set at 35°C. The sample was measured for 3h. The particle solution was removed from the sample cell via a syringe and centrifuged at 9000 rpm for 30 minutes. The supernatant was removed and the sedimented nanoparticles (NPs) re-dispersed in D$_2$O for further ex-situ TEM analysis.

### 8.3. Simultaneous SAXS/SANS setup:

SANS measurements were performed on beamline D22 at the ILL (Institut Laue–Langevin, Grenoble, France). Wavelength at λ = 0.6 nm, a collimation of 7 m and a sample-to-detector distance (SDD) of 8 m were used for the in-situ study. An acquisition time of 60 s was used at these conditions in synchronization with the SAXS measurements. The SAXS images were simultaneously acquired using the installed SAXS system at the D22 zone. By employing Cu *Kα* source with a photon flux of ≈ 1.1× 10$^7$ photons/s. For in situ simultaneous measurements, the sample-to-detector distance (SDD) was 1643 mm. All measurements were performed at 35 °C. The data have been corrected for sample thickness, acquisition time, and transmission. For 100-nm silica NPs, SANS measurements were performed at SSDs of 2.8, 8 and 17.6 m together with collimations of 2.5, 7 and 17.6 m, respectively.

### 8.4. SAXS and SANS data Analysis:

All SAXS measurements were for transmission, time and thickness corrected and absolute calibrated to 1 mm glassy carbon slab that was characterized at APL (Zhang *et al.*, 2010). The investigated nanostructures are assumed to exhibit a well-defined size and shape and have a homogeneous scattering length density. The scattered intensity displays the absolute scattering cross section $d\Sigma/d\Omega$ of a dispersion of NPs with interparticle interactions, which is described as (Schmutzler *et al.*, 2019):

$$\frac{d\Sigma}{d\Omega}(Q) = \frac{1}{V} <|F(Q)|^2> S(Q) ,$$

where $F(Q)$ is described as the form factor, $S(Q)$ is the structure factor and $Q$ the absolute of the scattering vector $\vec{Q} = |\vec{k_i} - \vec{k_s}|$. For elastic scattering the relation can be rewritten as $Q = 4\pi/\lambda \sin(\Theta)$. The previous parameters are corresponding to the incident wave vector $\vec{k_i}$, scattered wave vector $\vec{k_s}$, wavelength of the probe particle $\lambda$ and scattering angle $\Theta$.

From TEM images could be seen that there was a mixture of gold nanorods and strongly anisotropic particles inside the final solution. The scattering signal could be modeled best with a cylinder form factor by fitting length and radius. When neglecting interparticle interactions, the form factor is described as follows:

$$\frac{d\Sigma}{d\Omega}(Q) = \frac{volfrac}{V} \int_0^{\frac{\pi}{2}} F^2(R,L,Q,\alpha) \sin(\alpha)\, d\alpha + B ,$$

$$F(R,L,Q,\alpha) = 2(\Delta\rho)V \frac{\sin\left(\frac{1}{2}QL\cos(\alpha)\right)}{\frac{1}{2}QL\cos(\alpha)} \frac{J_1(QR\sin(\alpha))}{QR\sin(\alpha)}$$

whereas $R$ is the radius and $L$ the length of the cylinder, $\alpha$ describes the orientation of the cylinder to the Q-vector and orientationally averaged by the integral, $V$ is the volume of the cylinder $\pi R^2 L$, $\Delta\rho$ the scattering contrast between dispersion medium and particle, $J_1$ is the first order Bessel function, $volfrac$ is the volume fraction of particles in the observed volume and $B$ a constant describing the residual background. Furthermore, radius $R$ and length $L$ where distributed according to a gaussian function:

$$f(x,\mu,\sigma) = \frac{1}{Norm} \exp\left(-\frac{(x-\mu)^2}{2\sigma^2}\right) ,$$

$$\frac{d\Sigma}{d\Omega}(Q) = \frac{volfrac}{V} \iint_0^\infty \int_0^{\frac{\pi}{2}} f(x_R,R,\sigma_R) f(x_L,L,\sigma_L) F^2(x_R,x_L,Q,\alpha) \sin(\alpha)\, d\alpha\, dx_R\, dx_L + B ,$$

whereas $Norm$ equals a normalization factor, $\mu$ the mean of the distribution and $\sigma$ the standard deviation. The integrand was chosen to only contribute values of $\pm 3\sigma$ to decrease the load on the calculation. The radius and length were both fitted. To further constrain the scattering volume or volume fraction of particles, the last frame of the in-situ measurement was taken, and it was assumed that the number of particles per liter $N$ stays the same over the whole experiment. $N$ was calculated by dividing the volume fraction resulting from the fit of the final frame through the volume of one particle. The fit algorithm worked backwards from the final frame to the first one calculating the volume fraction corresponding to the fitted geometry of a single particle and the particle number per liter $N$. At around 35 minutes only background signal can be seen, resulting in unphysical values for radius and length, which is why those parameters are omitted in **Figure 9c** in the manuscript.

The SANS data was fitted by applying a well-defined model previously published by our group (Schmutzler *et al.*, 2018). The shape of the micelles can be described by using a core-shell ellipsoid model for the form factor of the scattered intensity:

$$F(Q,\alpha) = F_{core}(Q, R_{minor}, R_{major}, \Delta\rho_{core}, \alpha) + F_{shell}(Q, R_{minor}+t, R_{major}+t, \Delta\rho_{shell}, \alpha) ,$$

$$F(Q, R_1, R_2, \Delta\rho, \alpha) = \frac{3\Delta\rho V\, (\sin(Qr(R_1,R_2,\alpha)) - \cos(Qr(R_1,R_2,\alpha)))}{(Qr(R_1,R_2,\alpha))^3} ,$$

$$r(R_1, R_2, \alpha) = \sqrt{(R_1^2 \sin^2(\alpha) + R_2^2 \cos^2(\alpha))} ,$$

whereas $R_{minor}$ is the minor half axis, $R_{major}$ the major half axis and $t$ the thickness of the shell of the core-shell ellipsoid. Furthermore, $\Delta\rho_{shell}$ is corresponding to the scattering contrast between dispersion medium and shell, $\Delta\rho_{core}$ is corresponding to the scattering contrast between core and shell, $V$ is the volume of either the shell or core of the nanostructure and $\alpha$ is the angle between minor half axis and scattering vector $\vec{Q}$. To model the correlation peak between micelles a Hayter-Penfold structure factor was applied (Hayter & Penfold, 1981). Additionally, the shell thickness $t$ of the form factor was distributed by a gaussian function, which resulted in the following relation for the scattering cross section for randomly oriented particles and the corresponding structure factor $S(Q)$:

$$\frac{d\Sigma}{d\Omega}(Q) = \frac{1}{V} S(Q) \int_0^\infty \int_0^{\frac{\pi}{2}} f(x_t, t, \sigma_t) F^2(t, Q, \alpha) \sin(\alpha)\, d\alpha\, dx_t + B\ .$$

Here $V$ equals the volume of the core shell ellipsoid $V = \frac{4}{3}\pi(R_{major} + t)(R_{minor} + t)^2$ and $B$ the contribution of a constant background signal. The volume fraction, major half axis, charge of the micelles and salt concentration in the solution were fitted. As before the sequential fitting process started from the final frame to the first one. Predefined models from the package as models implemented in the program SasView (Doucet *et al.*, 2019) were used in combination with a custom-made python script to sequentially fit the in-situ SAXS and SANS data.

### 8.5. Transmission Electron Microscopy (TEM):

Microscopy images of the Au NPs synthesized in the experimental setup were carried out with a Philips CM30 transmission electron microscope. It was maintained at an acceleration voltage of 300 kV and utilized a LaB$_6$ cathode at a current of 24 mA. The samples were washed once more with H$_2$O, to reduce the amount of CTAB in the solution and drop-casted onto a copper grid covered with a thin carbon film (Plano).\

## 9. SAXS data fitting of Ludox NPs

Following a solvent (H$_2$O) subtraction, the SAXS profiles are modeled using SASfit (Bressler *et al.*, 2015) with only a sphere form factor with a log-norm distribution function. No structure factor is applied to the fitting. According to the fitting, SiO$_2$ NPs is bimodal distributed with a diameter of 26±2 nm and small fraction of aggregates of about 70 nm in size.

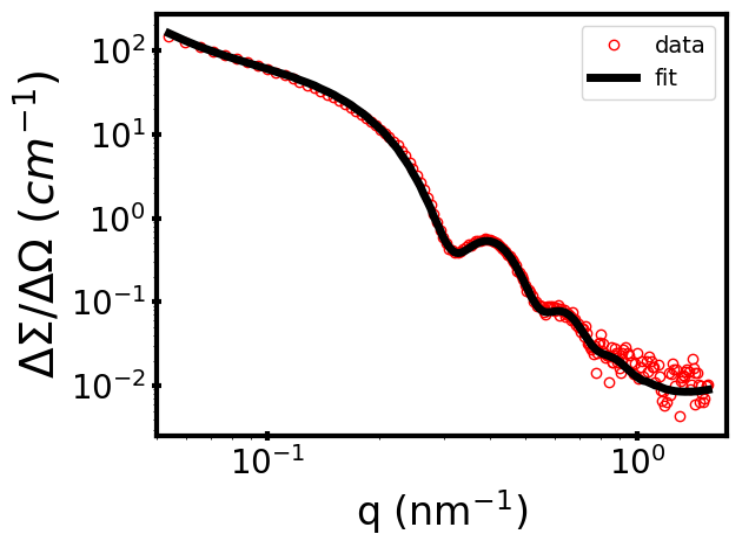

**Figure 1S.** 1D SAXS profiles (open circle) and the data fit (solid lines) of the Ludox TM50 (Si NPs) sample.

## 10. SAXS at two different sample angles

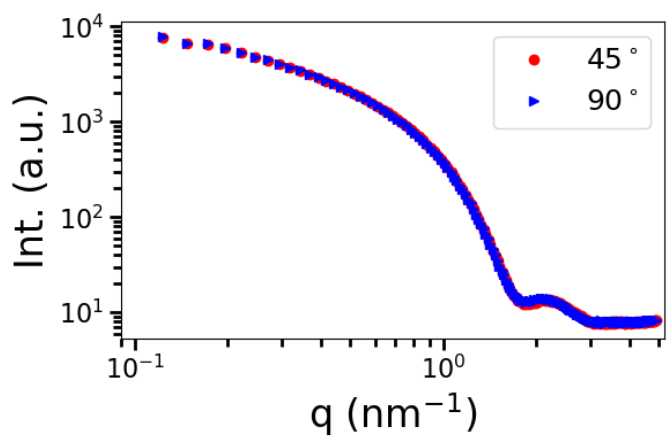

**Figure 2S.** 1D SAXS profiles of a CdSe nanoparticle sample. Correction of scattering profile, by multiplying each data point of the sample measured at 45° angle by 1.414, is perfectly coincidence with that collected at 90° angle.

**References:**

Bressler, I., Kohlbrecher, J. & Thunemann, A. F. (2015). *J Appl Crystallogr* **48**, 1587-1598.
Doucet, M., King, S., Butler, P., Kienzle, P., Parker, P., Krzywon, J., Jackson, A., Richter, T., Gonzales, M., Nielsen, T., Ferraz, L., R., Markvardsen, A., Heenan, R., Bakker, J. & Alina, G. (2019). *SasView version 5.0.0,* https://www.sasview.org.
Hayter, J. B. & Penfold, J. (1981). *Mol Phys* **42**, 109-118.
Schmutzler, T., Schindler, T., Goetz, K., Appavou, M. S., Lindner, P., Prevost, S. & Unruh, T. (2018). *J Phys-Condens Mat* **30**.
Schmutzler, T., Schindler, T., Zech, T., Lages, S., Thoma, M., Appavou, M.-S., Peukert, W., Spiecker, E. & Unruh, T. (2019). *ACS Applied Nano Materials* **2**, 3206-3219.
Vigderman, L. & Zubarev, E. R. (2013). *Chem Mater* **25**, 1450-1457.
Zhang, F., Ilavsky, J., Long, G. G., Quintana, J. P. G., Allen, A. J. & Jemian, P. R. (2010). *Metall Mater Trans A* **41a**, 1151-1158.

**Acknowledgements**

This work benefited from the use of the SasView application, originally developed under NSF award DMR-0520547. SasView contains code developed with funding from the European Union's Horizon 2020 research and innovation programme under the SINE2020 project, grant agreement No 654000.